# *p*-Hacking Inflates Type I Error Rates in the Error Statistical Approach but not in the Formal Inference Approach

*Mark Rubin*
*Durham University, UK*

May 21, 2026

## Abstract

*p*-hacking occurs when researchers conduct multiple significance tests (e.g., $p_1;H_{0,1}$ and $p_2;H_{0,2}$) and then selectively report tests that yield desirable (usually significant) results (e.g., $p_2 \leq 0.05;H_{0,2}$) without correcting for multiple testing (e.g., 0.05/2 = 0.025). In the present article, I consider *p*-hacking in the context of two philosophies of significance testing — the error statistical approach and the formal inference approach. I argue that although *p*-hacking inflates Type I error rates in the error statistical approach, it does not inflate them in the formal inference approach. Specifically, in the error statistical approach, the "actual" familywise error rate (e.g., $1 - [1 - 0.05]^2 = 0.098$ for two independent tests) is relevant because it covers both the reported and unreported tests in the "actual" test procedure (i.e., $p_1;H_{0,1}$ and $p_2;H_{0,2}$). In this approach, Type I error rate inflation occurs because the "actual" error rate (0.098) is higher than the nominal error rate (0.05). In contrast, in the formal inference approach, the "actual" familywise error rate is irrelevant because (a) the researcher does not report a statistical inference about the corresponding intersection null hypothesis (i.e., $H_{0,1} \cap H_{0,2}$), and (b) the "actual" familywise error rate does not license inferences about the reported individual hypotheses (i.e., $H_{0,2}$). Instead, in the formal inference approach, only the nominal error rate is relevant, and a comparison with the "actual" error rate is inappropriate. Implications for conceptualizing, demonstrating, and reducing *p*-hacking are discussed.

*Keywords*: error statistics; familywise error rate; null results; *p*-hacking; questionable research practices; Texas sharpshooter fallacy

*p*-hacking occurs when researchers conduct multiple significance tests and then selectively report tests that yield desired, usually significant, results without correcting for multiple testing. Hence, *p*-hacking represents a type of undisclosed cherry-picking or fishing for specific (significant) test results. Some common examples of *p*-hacking include selective outlier exclusion, selective covariate inclusion, outcome switching, and optional stopping or data peeking (Nagy et al., 2025; Stefan & Schönbrodt, 2023).

Although *p*-hacking is often described as a "questionable" research practice (John et al., 2012; e.g., Nagy et al., 2025; Reis & Friese, 2022), it is widely regarded as being statistically and ethically problematic (e.g., Miller et al., 2025; Pickett & Roche, 2018; Simmons et al., 2011; Stefan & Schönbrodt, 2023). It is also believed to be a major contributor to the replication crisis (e.g., Bishop, 2019; Nagy et al., 2025). In particular, *p*-hacking is thought to inflate Type I error rates above their nominal conventional level, resulting in a larger proportion of false positive results in the literature than would otherwise be expected (e.g., Lakens, 2019, p. 221; Nagy et al., 2025; Reis & Friese, 2022; Simmons et al., 2011; Stefan & Schönbrodt, 2023). This disproportionate number of false positives is then thought to cause unexpectedly low replication rates in studies that do not engage in *p*-hacking.

In the present article, I aim to add some nuance to this view by distinguishing between two philosophies of significance testing — the error statistical approach (Mayo, 1996, 2018) and the formal inference approach (Rubin, 2021b, 2024a, 2024b). I argue that, although *p*-hacking inflates Type I error rates in the error statistical approach, it does not inflate them in the formal inference approach. To be clear, I am not claiming that *p*-hacking does not exist, that it does not represent a bias, or that it does not inflate Type I error rates in some philosophies of significance testing. I am only claiming that *p*-hacking does not inflate Type I error rates in the formal inference approach to significance testing.

I should also clarify that, consistent with the error statistical approach, I am only concerned with classic frequentist conceptualizations of Type I error rates in the context of significance testing (e.g., Mayo, 2018, p. 187; Neyman & Pearson, 1928, p. 177). I am not concerned with alternative conceptualizations such as Bayesian error rates (e.g., Wagenmakers, 2018), false finding rates based on conditional posterior probabilities (Mayo, 2018, pp. 183–184; Mayo & Morey, 2017; Pollard & Richardson, 1987), the rejection of a false null hypothesis that is "closer to the truth" than a false alternative hypothesis (Dennis et al., 2019, p. 23), or false discovery rates (Benjamini & Hochberg, 1995).

The article proceeds as follows: Section 1 explains how *p*-hacking inflates Type I error rates in the error statistical approach. Sections 2 and 3 introduce the formal inference approach and explain why *p*-hacking does not inflate Type I error rates in this approach. Section 4 considers some key areas of agreement and disagreement between the two approaches. Section 5 considers the error statistical approach's minimal severity requirement and argues that it is not necessary to obtain valid Type I error rates in the formal inference approach. Section 6 considers the nondisclosure of null results during *p*-hacking and argues that, although this practice does not inflate Type I error rates in the formal inference approach, it may bias substantive inferences under certain circumstances. Finally, Section 7 summarizes the conclusions and discusses their implications for conceptualizing, demonstrating, and reducing *p*-hacking and for understanding and addressing the replication crisis.

# 1. *p*-Hacking Inflates Type I Error Rates in the Error Statistical Approach

## Defining Type I Error Rates

A Type I error rate represents the frequency with which a significance test would incorrectly reject a statistical null hypothesis during a hypothetical long run of repeated random sampling from a specified null population (Neyman & Pearson, 1928, pp. 176–177, 231–232). Without denying this Neyman-Pearson long-run "performance" rationale, error statisticians adopt a "probative" approach in which they consider the error-probing capacity of a test procedure in the single case at hand (Mayo, 2018, pp. 13–14, 162; Mayo & Spanos, 2006, p. 351; Mayo & Spanos, 2011, pp. 162–163). Here, a Type I error rate is used to "license" specific inferences based on the application of a particular test. As Mayo and Spanos (2006) explained, "pre-data, the choices for the type I and II errors reflect the goal of ensuring the test is capable of licensing given inferences severely" (p. 350; see also Mayo & Spanos, 2011, p. 167). For example, a test with a nominal Type I error rate of 0.05 ($\alpha$) is capable of licensing specific statistical inferences with a minimum "worst case" severity of 0.95 (i.e., $1 - \alpha$), assuming that the test has adequate statistical assumptions (Mayo, 1996, pp. 124, 304, 399, 407; Mayo, 2008, pp. 863–864; Mayo, 2018, p. 94; Spanos & Mayo, 2015, p. 3535).

## *p*-Hacking's Impact on Type I Error Rates

As Mayo and Spanos (2011, p. 166) explained, a key requirement in the error statistical approach is that the probability of observing a significant *p*-value should be $\leq \alpha$ when all null hypotheses are true (i.e., Pr[observing a *p*-value $\leq \alpha$] $\leq \alpha$). However, this probability becomes greater than $\alpha$ when multiple undisclosed significance tests are performed in the experimental testing context. Hence, when *p*-hacking occurs, the probability of making at least one Type I error becomes greater than $\alpha$.

Consequently, a test procedure that includes *p*-hacking will have an "actual" Type I error rate that is higher than the nominal "computed" error rate of $\alpha$ (Mayo, 1996, pp. 303–304, 348; Mayo, 2008, pp. 874–875; Mayo, 2018, pp. 274–275; Mayo & Cox, 2010, pp. 267–270). It is this discrepancy between "actual" and nominal error rates that is conceptualized as Type I error rate "inflation" (Mayo, 2025, p. 1).

To illustrate, imagine that a researcher conducts significance tests on 20 null hypotheses, $H_{0,1}$, $H_{0,2}$, $H_{0,3}$, … $H_{20}$, yielding 20 corresponding *p*-values, $p_1$, $p_2$, $p_3$, … $p_{20}$. They then search for results that are significant at the nominal conventional level of 0.05. They find that only $p_{13}$;$H_{0,13}$ is significant, and so they selectively report that particular result.[1]

In this situation, the "actual" test procedure includes searching for significance among 20 tests and then selectively reporting significant results (Mayo, 1996, p. 304). Consequently, the "actual" Type I error rate is the familywise error rate for those 20 tests (Mayo, 2018, p. 275). The upper bound of the familywise error rate can be computed using the formula $1 - (1 - \alpha)^k$, where $k$ is the number of independent significance tests in the "actual" test procedure. Hence, the upper bound of the "actual" error rate is 0.642 (i.e., $1 - [1 - 0.05]^{20}$).[2] In contrast, the nominal error rate for the reported inference about $H_{0,13}$ is 0.05. Hence, the "actual" error rate is inflated above the nominal error rate, and the nominal error rate is "invalidated" or "vitiated" because it does not match the "actual" error rate (i.e., Pr[observing a *p*-value $\leq \alpha$] $> \alpha$; Mayo, 2008, p. 876; Mayo, 2018, pp. 234, 285, 438; Mayo & Spanos, 2011, p. 190).

To address this potential problem, error statisticians recommend that we check or "audit" the experimental testing context for any "biasing selection effects," such as *p*-hacking (e.g., by cross-referencing with a preregistered analysis plan; Mayo, 2018, pp. 92, 106–107, 275, 439). The "minimal severity requirement" is that we attempt to rule out ways in which a claim may be false (Mayo, 2018, p. 5). Biasing selection effects occur when hypotheses and data are generated and selected in ways that violate, secretly alter, or prevent an assessment of this requirement (Mayo, 2018, pp. 92, 269). Hence, *p*-hacking represents a biasing selection effect because its hidden multiple testing secretly inflates the "actual" Type I error rate and prevents us from effectively ruling out random sampling error as an explanation for the observed result. Consequently, we are left with "bad evidence, no test (BENT)" (Mayo, 2018, p. 5).

# 2. The Formal Inference Approach

The error statistical approach can be contrasted with what I call here the formal inference approach (Rubin, 2021a, 2021b, 2024a, 2024b, 2025). In this approach, relevant Type I error rates are based on formally reported (publicly specified) statistical inferences rather than "actual" test procedures, which may not be reported.[3] In particular, the contents of a formally reported inference are used to determine (a) the category of statistical inference, (b) the number of significance tests associated with the inference, and (c) the specified alpha level for each test (or the conventional alpha level if none is specified). This information is then used to compute the relevant Type I error rate for the inference (e.g., Rubin, 2024b, p. 52).

## Categories of Statistical Inferences

In the formal inference approach, a statistical inference consists of (a) a provisional decision to either reject or fail to reject a formally specified statistical null hypothesis and (b) the result of one or more significance tests that indicate conditional uncertainty about this decision (Fisher, 1956, pp. 57, 108–110; Rubin, 2024b, p. 49). Three categories of statistical inference are distinguished: *individual*, *union-intersection*, and *intersection-union* (Berger, 1982; García-Pérez, 2023; Hochberg & Tamhane, 1987; Parker & Weir, 2020; Roy, 1953; Rubin, 2021b, 2024a, 2024b).

### *(1) Individual Inference*

If a researcher makes a decision about an *individual* alternative hypothesis (e.g., $H_{1,1}$) and corresponding *individual* null hypothesis (e.g., $H_{0,1}$) using a nominal alpha level of α, then the Type I error rate for that decision is α, even if the researcher makes multiple such decisions (reported or unreported) within the same study. In this case, the individual null hypothesis can only be rejected following a significant result on *one* particular test (i.e., $p_1 \leq \alpha; H_{0,1}$). Hence, there is only a *single* opportunity to make a Type I error with regard to that particular individual null hypothesis (see also García-Pérez, 2023, p. 15; Parker & Weir, 2020, pp. 563–564; Rubin, 2017a, pp. 271–272; Rubin, 2020, p. 380; Rubin, 2021a; Rubin, 2021b, pp. 10978–10983; Rubin, 2024a, p. 3; Rubin, 2024b, p. 51).

### *(2) Union-Intersection Inference*

If a researcher makes a decision about a *union* alternative hypothesis (e.g., $H_{1,1} \cup H_{1,2} \cup H_{1,3} \cup \ldots H_{1,k}$) and corresponding *intersection* null hypothesis ($H_{0,1} \cap H_{0,2} \cap H_{0,3} \cap \ldots H_{0,k}$), then the Type I error rate for that decision is the familywise error rate of $1 - (1 - \alpha)^k$. In this case, the intersection null hypothesis can be rejected following *at least one* significant result (i.e., $p_{\min} \leq \alpha$) among *any* of the *k* tests ($p_1;H_{0,1}$, $p_2;H_{0,2}$, $p_3;H_{0,3}$, … $p_k;H_{0,k}$; e.g., García-Pérez, 2023, p. 15; Hochberg & Tamhane, 1987, p. 28; Parker & Weir, 2020, p. 563; Roy, 1953; Rubin, 2021b, pp. 10973–10975; Rubin, 2024b, pp. 51–52). Hence, there are *multiple* (*k*) opportunities to make a Type I error with regard to the intersection null hypothesis. Consequently, the familywise error rate for a decision about the intersection null hypothesis ($\alpha_{\text{Intersection}}$) is higher than the nominal error rate for tests of each of its constituent null hypotheses ($\alpha_{\text{Constituent}}$). To control $\alpha_{\text{Intersection}}$ at a conventional level (e.g., 0.05), $\alpha_{\text{Constituent}}$ must be adjusted below the conventional level (e.g., using a Dunn–Šidák correction, $\alpha_{\text{Constituent}} = 1 - [1 - \alpha_{\text{Intersection}}]^{1/k}$, or a Bonferroni correction, $\alpha_{\text{Constituent}} = \alpha_{\text{Intersection}}/k$).[4]

***(3) Intersection-Union Inference***

If a researcher makes a decision about an *intersection* alternative hypothesis (e.g., $H_{1,1} \cap H_{1,2} \cap H_{1,3} \cap \ldots H_{1,k}$) and corresponding *union* null hypothesis ($H_{0,1} \cup H_{0,2} \cup H_{0,3} \cup \ldots H_{0,k}$), then the Type I error rate for that decision is the nominal α level. Here, the union null hypothesis can only be rejected following *all* significant results (i.e., $p_{\max} \leq \alpha$) on *all k* tests ($p_1;H_{0,1}$, $p_2;H_{0,2}$, $p_3;H_{0,3}$, … $p_k;H_{0,k}$). The alpha level for each constituent test ($\alpha_{\text{Constituent}}$) does not need to be reduced in this case because the researcher only has a *single* opportunity to erroneously reject the union null hypothesis (Berger, 1982, p. 295; Rubin, 2021b, p. 10976). Indeed, $\alpha_{\text{Constituent}}$ may be *increased* to address the low statistical power of the intersection-union test (Francis & Thunell, 2021).

## Guiding Principles for Determining Type I Error Rates

The formal inference approach applies the following four guiding principles when determining relevant Type I error rates.

***(1) Type I Error Rates Refer to Random Error, not Systematic Error***

A Type I error rate is represented by one or more critical regions in the tail areas of the specified sampling distribution of a test statistic given under a statistical null hypothesis (Mayo, 2018, p. 187; Neyman, 1950, p. 265). Hence, it represents the probability of incorrectly rejecting a null hypothesis due to random sampling error *given* a specified sampling distribution. It is not supposed to quantify the probability of incorrectly rejecting a null hypothesis due to systematic error (Meehl, 1997, pp. 397, 401; Neyman & Pearson, 1928, p. 176; Pollard & Richardson, 1987, p. 162; Rubin, 2024a, p. 3, Confusion IV; Rubin, 2024b, p. 49). Systematic error can include (a) theoretical error (e.g., errors in theories, hypotheses, predictions, background knowledge, causal relations, constructs, etc.); (b) methodological error (e.g., errors in research design, sampling procedures, testing conditions, stimuli, manipulations, measures, controls, etc.); (c) data error (e.g., errors in data collection, entry, coding, cleaning, aggregation, transformation, etc.); or (d) analytical error (e.g., errors in statistical assumptions such as normality, linearity, or independence). A Type I error rate does not represent any of these types of systematic error. Instead, it represents a hypothetical, mathematical, subjunctive conditional probability that assumes that all test assumptions have been met, there is no systematic error, and the only remaining source of decision-making error is random sampling error, as represented by the critical tail area(s) of the sampling distribution (Cortina & Dunlap, 1997, pp. 166–167; Fisher, 1956, p. 44; Fisher, 1959, pp. 21–22; Kass, 2011, pp. 5–6, 7; Meehl, 1990, p. 111; Neyman, 1952, pp. 23, 57; Neyman, 1955, p. 17; Neyman & Pearson, 1928, pp. 176–177, 232; Rubin, 2024b, pp. 47–48).

The distinction between random and systematic error is consistent with the distinction between *statistical* and *substantive* (nonstatistical) hypotheses (Meehl, 1990, p. 116; 1997, p. 401;

Neyman, 1950, p. 290; Rubin, 2024b, p. 49). A statistical hypothesis is a mathematical statement about a random variable in a particular study (e.g., $H_0$: $\mu_1 = \mu_2$; Neyman, 1950, p. 290). In contrast, a substantive hypothesis is a nonstatistical statement about a theoretical effect that is assumed to generalize beyond a particular study (e.g., "men have higher self-esteem than women"). Importantly, a Type I error rate only refers to the incorrect rejection of a *statistical* null hypothesis due to random sampling error. It does not represent the probability of incorrectly rejecting a *substantive* null hypothesis, because this decision may also be influenced by systematic error in theory, method, data, and/or analysis, and Type I error rates do not index systematic error. Indeed, the probability of incorrectly rejecting a *substantive* null hypothesis may be considerably greater than the Type I error rate because it is based on not only random sampling error but also systematic error (Rubin, 2024b, p. 50).

Furthermore, systematic error can cause the substantive misinterpretation of a significant statistical result. For example, a theoretically uninteresting "crud" effect may be misinterpreted as a theoretically important effect (Meehl, 1997, p. 402).[5] Again, it is crucial not to conflate this substantive interpretational error with a statistical Type I error (Greenland, 2017, p. 639; Meehl, 1990, p. 125; Meehl, 1997, p. 402; Rubin, 2017a, p. 274; Rubin, 2021b, p. 10986; Rubin, 2024a, p. 3, Confusion IV; Rubin, 2024b, pp. 50–51). The former is a substantive misinterpretation of a true positive result caused by systematic error. The latter is a statistical misclassification of a true negative (null) result caused by random error.

***(2) Type I Error Rates for Different Tests are Incommensurate with One Another***

A Type I error rate is a hypothetical probability that is contingent on a specified sampling distribution under a statistical null hypothesis. It cannot be detached from its sampling distribution and treated as a free-floating unconditional probability that may become "inflated," "altered," "invalidated," or "vitiated" when moving from one test to another (Cortina & Dunlap, 1997, pp. 166–167; Pollard & Richardson, 1987; Weisberg, 2014, p. 289).

For example, the Type I error rate for one brand of pregnancy test may be lower than that for another brand. However, it would be inappropriate to claim that the second test's error rate is an "inflated" version of the first test's error rate or that the first test's error rate is "invalidated" in this context. Each error rate licenses a separate decision about a different statistical null hypothesis based on a different statistical model. In this respect, Type I error rates for different tests are model-contingent probabilities that are independent and incommensurate with one another (García-Pérez, 2023, p. 15; Rubin, 2024a, p. 3).

***(3) Familywise Error Rates License Union-Intersection Inferences, not Individual Inferences***

The familywise error rate for the intersection null hypothesis $H_{0,1} \cap H_{0,2}$ only licenses a *union-intersection* inference about $H_{0,1} \cap H_{0,2}$ as a whole. In this case, the constituent null hypotheses $H_{0,1}$ and $H_{0,2}$ are treated as being logically exchangeable with one another rather than as distinct individual hypotheses (García-Pérez, 2023, p. 2; Rubin, 2021b, pp. 10978–10982; Rubin, 2024a, p. 2). Consequently, during union-intersection testing, a significant result in relation to the constituent null hypothesis $H_{0,1}$ does not allow us to reject $H_{0,1}$ as an individual hypothesis. It only allows us to reject $H_{0,1} \cap H_{0,2}$ and conclude that either $H_{0,1}$ or $H_{0,2}$ or both are false.[6]

The same principle applies in the case of a one-way ANOVA (García-Pérez, 2023, p. 6). A significant omnibus result allows us to claim that there is a significant difference between at least one pair of means, but it does not allow us to identify which pair (Rubin, 2024a, p. 2).

Hence, in the formal inference approach, a familywise error rate is not the probability of making "at least one Type I error" because this phrasing implies that the researcher is making multiple *individual* inferences (decisions) about multiple *individual* hypotheses, each licensed by its own nominal Type I error rate (Rubin, 2021b, pp. 10978–10983; Rubin, 2024a, p. 3, Confusion III; Rubin, 2024b, p. 59). Instead, as its name implies, the familywise error rate represents the probability of making an error about an entire *family* of hypotheses, not any specific individual hypothesis within that family.

Following a union-intersection test, the researcher makes a *single* decision to either reject or fail to reject an intersection null hypothesis (e.g., $H_{0,1} \cap H_{0,2}$) on the basis of "at least one" significant result among their constituent tests (e.g., $p_1;H_{0,1}$ and $p_2;H_{0,2}$; see also García-Pérez, 2023, p. 5; Rubin, 2021b, p. 10982). The familywise error rate is then the probability of making a Type I error about this intersection null hypothesis rather than the probability of making at least one Type I error about multiple individual null hypotheses.

***(4) Type I Error Rates are Based on Formally Reported Inferences, not "Actual" Test Procedures***

If a researcher's "actual" test procedure includes tests of $H_{0,1}$ and $H_{0,2}$, then it is possible for them to report statistical inferences about $H_{0,1}$, $H_{0,2}$, $H_{0,1} \cap H_{0,2}$, and/or $H_{0,1} \cup H_{0,2}$. Nothing in the description of the test procedure (i.e., "tests of $H_{0,1}$ and $H_{0,2}$") determines which of these inferences will be reported. Of course, a researcher can privately intend and publicly plan to report specific inferences. However, intentions and plans can change and, when they do, inferences and associated error rates may change with them (for related points, see Weisberg, 2014, p. 289). For example, a researcher may plan to make an inference about $H_{0,1} \cap H_{0,2}$, which would require an alpha adjustment. However, they may then change their mind and decide to make two individual inferences instead, one about $H_{0,1}$ and one about $H_{0,2}$, neither of which would require an alpha adjustment. Hence, from a formal inference perspective, it is formally reported statistical inferences, rather than "actual" test procedures and/or planned inferences, that determine relevant Type I error rates (Rubin, 2024b. pp. 59, 62). In particular, a formal statistical inference serves as the basis for the "rational reconstruction" of a hypothetical test procedure and its subjunctive conditional error rate (Reichenbach, 1938, pp. 6–7). For example, an individual inference about $H_{0,2}$ would be associated with a hypothetical test procedure that only refers to $H_{0,2}$ and not $H_{0,1}$.

## Conceptual Roots and Relations

The formal inference approach is latent within the writings of Fisher (1956), Popper (1966, 2002), and others (e.g., Hennig, 2010, 2023; Kass, 2011). In particular, it fits Fisher's (1956) view that "the only populations that can be referred to in a test of significance have no objective reality, being exclusively the product of the statistician's imagination through the hypotheses which … [they have] decided to test" (Fisher, 1956, p. 77). Consequently, significance tests "are based on hypothetical probabilities calculated from their null hypotheses. They do not generally lead to any probability statements about the real world" (Fisher, 1956, p. 44). From this perspective, Type I error rates are contingent on *hypothetical* sampling distributions given under *formally specified* hypotheses, rather than "*actual*" sampling distributions based on *potentially unreported* "actual" test procedures (e.g., Cortina & Dunlap, 1997, pp. 166–167).

The formal inference approach also fits with more contemporary views of significance testing. For example, it is consistent with Greenland's (2017) view that "statistical analyses are merely thought experiments, informing us as to what would follow deductively under their assumptions" (p. 640; see also Kass, 2011, p. 7; Weisberg, 2014, p. 289) and Hennig's (2010, 2023) frequentism-as-model approach, which assumes that "mathematical models are thought constructs and means of communication about reality" (Hennig, 2023, p. 16; see also Zhao, 2025).

The formal inference approach also follows some classic views in the philosophy of science. In particular, its focus on formally reported inferences and hypotheses follows Popper's methodological rule that scientific hypotheses must be "intersubjectively testable" so that, in principle, they can be tested by anyone (Popper, 1966, p. 218; Popper, 2002, p. 22). The error statistical approach faces problems meeting Popper's requirement because, during *p*-hacking, the "actual" test procedure of multiple testing is undisclosed, and so it cannot be accessed and repeated by other researchers.

More broadly, the formal inference approach is consistent with Popper's (1974, 1983, 1994, 2002) theory-centric deductive method of testing and his critical rationalist approach to theory appraisal (Rubin, 2025). In contrast, the error statistical approach follows Peirce's (1958) quasi-experimental method (Mayo, 1996, p. 431), and it is consistent with the New Experimentalist view that "experiments, as Ian Hacking taught us, live lives of their own, apart from high level theorizing" (Mayo, 1996, p. xiii). Consequently, the formal inference approach is better suited for falsifying (logically refuting rather than completely rejecting) universal hypotheses and theories (e.g., "all swans are white"; Rubin, 2025, p. 8; see also Popper, 2002, pp. 47–48), whereas the error statistical approach is better suited for establishing the existence of local experimental effects (e.g., "all swans tested using this method under these conditions are white"; Mayo, 1996, pp. xiii, 12, 213; Mayo, 2018, pp. 94–95; Rubin, 2025, p. 8). In addition, as I explain later, the formal inference approach is more applicable in the case of unplanned exploratory analyses (Rubin, 2017a, 2020, 2022, 2024b; Rubin & Donkin, 2024).

# 3. *p*-Hacking Does Not Inflate Type I Error Rates in the Formal Inference Approach

## The Basic Argument

To recap, the main claim is that *p*-hacking inflates Type I error rates in the error statistical approach but not in the formal inference approach. The basic argument for this claim is as follows.

In the error statistical approach, the "actual" familywise error rate is relevant during *p*-hacking because it covers both reported and unreported tests in the "actual" test procedure. In this approach, Type I error rate inflation occurs because the "actual" familywise error rate is higher than the nominal rate for the reported inference.

In contrast, in the formal inference approach, the "actual" familywise error rate is irrelevant because it only licenses an inference about an intersection null hypothesis (e.g., $H_{0,1} \cap H_{0,2}$), and, in practice, researchers who *p*-hack cannot report this inference because they do not disclose some of their tests (e.g., $p_1 > 0.05;H_{0,1}$). Instead, they usually report separate individual inferences about selectively reported individual null hypotheses (e.g., $p_2 \leq 0.05;H_{0,2}$).[7] Logically, the "actual" familywise error rate does not license these individual inferences. Rather, they are licensed by their own separate nominal error rates. *p*-hacking does not inflate these nominal error rates because inflation only occurs relative to the "actual" familywise error rate, and a comparison with the "actual" error rate is inappropriate because it is incommensurate and inconsistent with the reported individual inferences.

## An Illustration

To illustrate, imagine the following example of *p*-hacking: A researcher conducts a two-sided independent samples *t* test using a conventional alpha level of 0.05. They fail to find a significant result with regard to their first null hypothesis $H_{0,1}$: $t(326) = 1.88$, $p = 0.061$. *For that reason*, they remove an outlier from their sample and conduct a second test. This time, the researcher finds a significant result, $t(325) = 2.16$, $p = 0.032$, which they then report without disclosing the nonsignificant result from their first test.

Note that, even if the researcher is testing the same *substantive* null hypothesis in this scenario (e.g., "there is no gender difference in self-esteem"), they are testing two different *statistical* null hypotheses (Meehl, 1990, p. 125; Meehl, 1997, p. 401; Neyman, 1950, p. 290). In particular, the second test has a smaller degrees of freedom than the first, and it is formally specified by a different test procedure (i.e., outliers removed). Hence, it tests a different statistical null hypothesis, $H_{0,2}$, with a different statistical model that includes a slightly different sampling distribution and sample space.

From an error statistical perspective, the researcher has conducted two significance tests ($p_1$;$H_{0,1}$ and $p_2$;$H_{0,2}$) and then selectively reported whichever test(s) yield a significant result (Mayo, 1996, pp. 303–304, 348; Mayo, 2008, pp. 874–875; Mayo, 2018, pp. 274–275; Mayo & Cox, 2010, pp. 267–270). In this case, the "actual" sampling distribution is given under the "actual" test procedure's "global" or "universal" intersection null hypothesis, $H_{0,1} \cap H_{0,2}$ (Mayo, 2008, p. 875; Mayo, 2018, p. 276; Mayo & Cox, 2010, p. 269). Consequently, the "actual" error rate is the familywise error rate 0.098 (i.e., $1 - [1 - 0.05]^2$), which is inflated above the nominal error rate of 0.05.

In contrast, from a formal inference perspective, the relevant sampling distribution is given under $H_{0,2}$, not $H_{0,1} \cap H_{0,2}$, because the researcher only reported an *individual* inference about $H_{0,2}$, not a *union-intersection* inference about $H_{0,1} \cap H_{0,2}$. Consequently, the relevant test procedure is the formally reported procedure for $H_{0,2}$, in which any outliers are removed, and the relevant Type I error rate is the nominal error rate (Rubin, 2020, p. 381; Rubin, 2021b, p. 10991; Rubin, 2025, p. 10).

Importantly, from a formal inference perspective, it is not appropriate to compare the "actual" familywise error rate for a decision about $H_{0,1} \cap H_{0,2}$ with the nominal error rate for a decision about $H_{0,2}$ and argue that the former represents an "inflated" version of the latter. As discussed previously, these two error rates are incommensurate with one another because they refer to two separate decisions about two different statistical null hypotheses based on two different statistical models. Arguing that a test's Type I error rate has been inflated because another test has a larger error rate is like arguing that a person's height has been inflated because their friend is taller than them!

It is also logically inconsistent to use a familywise error rate to license an individual inference. We would be making a *fallacy of division* or *ecological fallacy* in this case because we would be misapplying an aggregate-level union probability to an individual member of the aggregate (Rubin, 2024a, p. 4; Selvin, 1958; Waller, 2018). In particular, the 0.098 union probability of obtaining *at least one* significant result given $H_{0,1} \cap H_{0,2}$ (i.e., the familywise error rate) does not represent the individual probability of obtaining a significant result given $H_{0,2}$, which remains at the nominal level of 0.05 (García-Pérez, 2023; Rubin, 2017a, pp. 271–272; Rubin, 2020, p. 380; Rubin, 2021a; Rubin, 2021b, pp. 10978–10983; Rubin, 2024a, p. 3; Rubin, 2024b, p. 51).

The familywise error rate would only be relevant if $H_{0,2}$ was treated as a logically exchangeable constituent of $H_{0,1} \cap H_{0,2}$, rather than as a distinct individual hypothesis. In this case, the researcher would report a union-intersection inference about $H_{0,1} \cap H_{0,2}$ as a whole, rather than an individual inference about $H_{0,2}$ alone (García-Pérez, 2023, p. 2; Rubin, 2021b, p. 10981; Rubin, 2024a, p. 2). In particular, following at least one significant result among $p_1$;$H_{0,1}$ and $p_2$;$H_{0,2}$, the union-intersection inference would be that "either $H_{0,1}$ or $H_{0,2}$ or both are false," rather than "$H_{0,2}$ is false."

In summary, from an error statistical perspective, (a) the "actual" test procedure includes two significance tests: $p_1$;$H_{0,1}$ and $p_2$;$H_{0,2}$; (b) the "actual" sampling distribution is given under the intersection null hypothesis $H_{0,1} \cap H_{0,2}$; and so (c) the "actual" Type I error rate is the familywise error rate of 0.098 (i.e., $1 - [1 - 0.05]^2$), which is "inflated" relative to the nominal error rate of 0.05. In contrast, from a formal inference perspective, (a) the formally reported inference is an individual inference about the individual null hypothesis $H_{0,2}$; (b) the relevant sampling distribution for this inference is given under $H_{0,2}$; and so (c) the relevant Type I error rate is $p_2$;$H_{0,2}$'s nominal error rate of 0.05, which is incommensurate with the familywise error rate for a decision about $H_{0,1} \cap H_{0,2}$. Figure 1 illustrates these differences.[8]

**Figure 1**

*p-Hacking in the Formal Inference and Error Statistical Approaches*

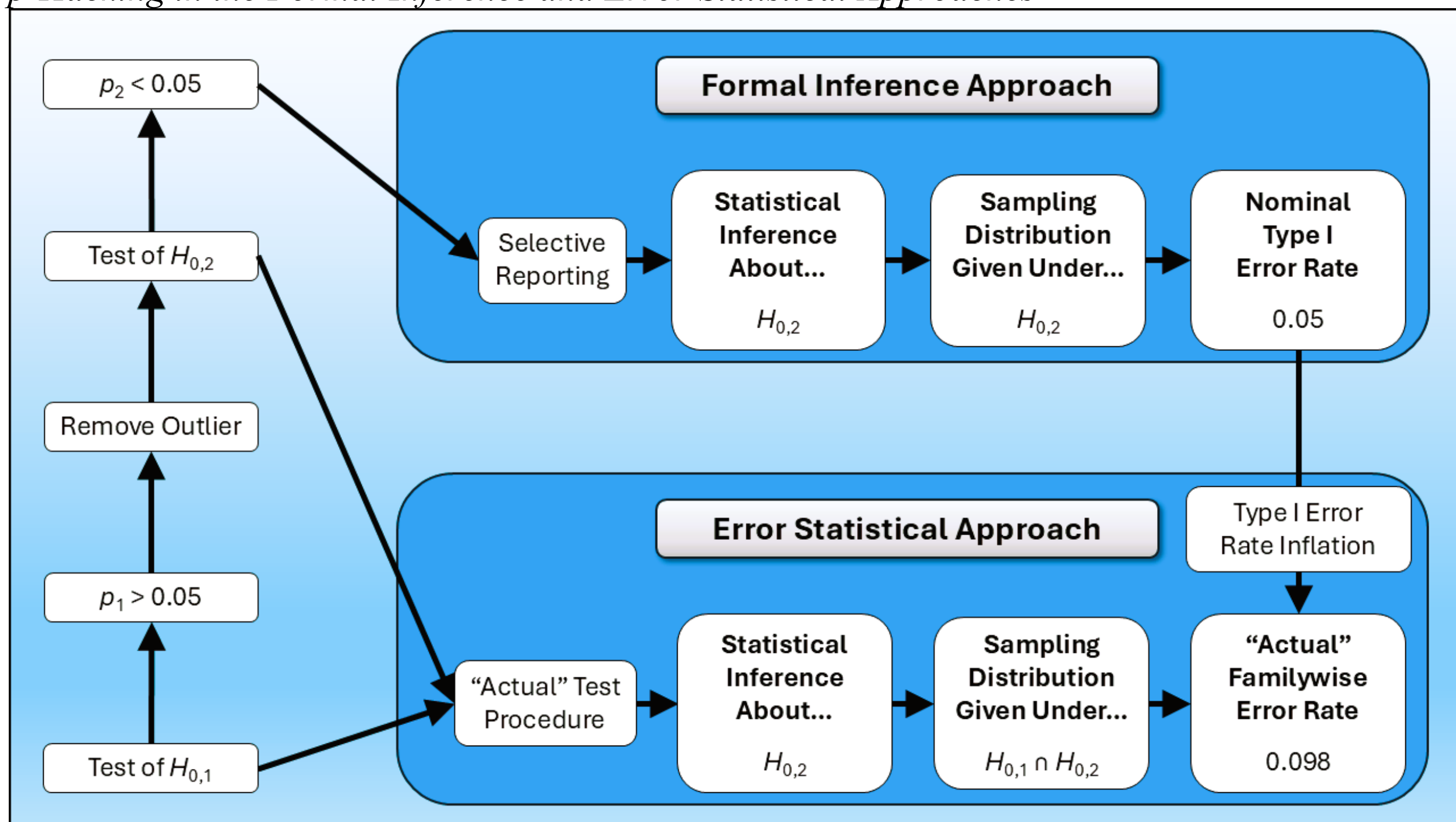

## A Closer Look

It is worth taking a closer look at the situation in Figure 1 in order to unpack some of its complexities. Contrary to the formal inference approach, the error statistical approach argues that the "actual" familywise error rate is relevant because we need to consider all of the hypotheses that could have been outputted by the "actual" test procedure (i.e., $H_{0,1}$ and $H_{0,2}$) "in order to assess the well-testedness of the one hypothesis [that it] … happened to have outputted" (i.e., $H_{0,2}$; Mayo, 2008, p. 876). As Mayo (2008) explained:

> The hypothetical error rates teach us about the test's capacities in the case at hand. Because at least one such impressive departure is common even if all are due to chance, the 'hunting expedition' has scarcely reassured us that it has done a good job of avoiding such a mistake in *this* case. The .05 'computed' *p*-value is invalidated when it comes to the 'actual' value. (p. 876, italics in original here and in all subsequent quotations)

Again, however, from a formal inference perspective, this reasoning commits the fallacy of division because it confounds union and individual probabilities. In the situation shown in Figure 1, the "actual" familywise error rate (a union probability) only informs us about the capacity of the "hunting expedition" to avoid an error in relation to the *intersection* null hypothesis $H_{0,1} \cap H_{0,2}$. It does not inform us about an *individual* test's capacity to avoid an error in relation to the *individual* null hypothesis $H_{0,2}$ (an individual probability), and it is $H_{0,2}$ that represents "the case at hand" in Figure 1, not $H_{0,1} \cap H_{0,2}$. From this perspective, it is the individual test's nominal error rate that indicates its error-probing capacity with respect to $H_{0,2}$, and that capacity is not invalidated by the hunting expedition that was used to identify and selectively report the test. By analogy, if 20 athletes are tested for illicit drugs, then hunting for, finding, and selectively reporting a test that yields a positive result does not alter that test's capacity to avoid an error in relation to the athlete who took the test (see also Rubin, 2024b, p. 55).[9]

Error statisticians argue that we "ought" to audit the experimental testing context to try to detect any *p*-hacking and make the "actual" test procedure objective (e.g., Mayo, 2018, p. 49). However, from a formal inference perspective, publicly reporting previously undisclosed tests does not necessarily change the specified category of statistical inference. For example, a researcher might report both $p_1;H_{0,1}$ and $p_2;H_{0,2}$ but specify two *individual* inferences as follows:

> One outlier was identified at ±3.00 *SD*s from the sample mean. There was no significant result when this outlier was included in the analysis, $t(326) = 1.88$, $p = 0.061$. However, there was a significant result when the outlier was excluded, $t(325) = 2.16$, $p = 0.032$.

In this case, the Type I error rate for each individual statistical inference remains at its nominal level even if the second test was "actually" conducted *because* the first test yielded a null result (see also García-Pérez, 2023; Parker & Weir, 2020, pp. 563–564; Rubin, 2017a, pp. 271–272; Rubin, 2020, p. 380; Rubin, 2021a; Rubin, 2021b, pp. 10978–10979; Rubin, 2024a, p. 3, Confusion I; Rubin, 2024b, p. 51). It would be a category mistake to use the familywise error rate to license individual inferences about $H_{0,1}$ and $H_{0,2}$, even if the result for $H_{0,1}$ inspired the test of $H_{0,2}$ (Rubin & Donkin, 2024, pp. 2023–2024).

In theory, the researcher could formally report a union-intersection inference based on the "actual" intersection null hypothesis $H_{0,1} \cap H_{0,2}$. For example, given different results, they could report that "there was a significant result when either including outliers, excluding outliers, or in both cases, $p_{min} \leq 0.025$." In this case, the formal inference and error statistical approaches would agree that the familywise error rate is relevant because, logically, the relevant sampling distribution for this formally reported union-intersection inference is given under $H_{0,1} \cap H_{0,2}$ (Rubin, 2017b, p. 323; Rubin, 2021b, pp. 10973–10975; Rubin, 2024b, pp. 51–52). However, in practice, the researcher cannot report this inference because their *p*-hacking conceals the null result for $H_{0,1}$.

Mayo (1996) argued that the nominal error rate misrepresents "*what should be expected to happen* … in subsequent experiments" (p. 349, see also p. 310). This "error unreliability" argument assumes that the original experiment "actually" undertook a union-intersection test of $H_{0,1} \cap H_{0,2}$, rather than an individual test of $H_{0,2}$, and then evaluated the test result using the wrong (nominal) sampling distribution (Spanos & Mayo, 2015, p. 3546). However, in the formal inference approach, a significance test of $H_{0,2}$ can be treated as either (a) part of a union-intersection test of $H_{0,1} \cap H_{0,2}$ or (b) an individual test of the individual null hypothesis $H_{0,2}$, and it is the formally reported inference that determines which is the case. Hence, if an original experiment reports an individual inference about $H_{0,2}$, then, objectively, the significance test is treated as an individual test of $H_{0,2}$, and "what should be expected to happen" in hypothetical exact replications of this individual test is licensed by that test's nominal Type I error rate (see also Rubin, 2024b, p. 58).

Finally, it is important to appreciate that the error statistical scenario shown in the bottom half of Figure 1 represents a simplistic conceptualization of the "actual" test procedure because it ignores the indeterminate nature of *p*-hacking in a hypothetical long run of repeated random sampling. By definition, a researcher who *p*-hacks will continue testing until they obtain a desired result. Consequently, although the researcher in Figure 1 managed to obtain a significant result on their second attempt in the current sample, they may need many more attempts in other samples (e.g., via selective covariate inclusion, outcome switching, etc.; Stefan & Schönbrodt, 2023). In this case, the number of tests ($k$) in hypothetical repetitions of the "actual" test procedure is unknown, and so we cannot compute the "actual" familywise error rate as $1 - (1 - \alpha)^k$ (Hochberg & Tamhane, 1987, p. 6; Mayo, 1996, pp. 313–314; Rubin, 2017b, p. 325; Rubin, 2021b, p. 10992; Rubin, 2024b, p. 56; see also Simmons et al., 2011, p. 1365).

From an error statistical perspective, this indeterminate testing problem invalidates significance testing during not only secretive *p*-hacking but also transparently reported result-dependent exploratory analyses (Rubin, 2017a; Rubin, 2021b, p. 10992; Rubin, 2024b, p. 56). Even so-called "honest hunting" for significance, which includes a correction for multiple testing, is untenable in an open-ended exploratory situation because transparency about the number of tests conducted in relation to one sample does not inform us about the number of tests that would be conducted in other samples (Mayo, 1996, pp. 311−314; Mayo, 2018, p. 275).

The formal inference approach avoids this problem by defining $k$ relative to a closed, formally specified inference rather than an open-ended and indeterminate "actual" test procedure (Rubin, 2024b, p. 64). For example, $k = 1$ for an individual inference about $H_{0,1}$, and $k = 2$ for a union-intersection inference about $H_{0,1} \cap H_{0,2}$, even if these two inferences have been selectively reported from a larger set of unreported inferences. Hence, from a formal inference perspective, significance testing retains its value in unplanned exploratory analyses (Rubin, 2017a, 2020, 2022, 2024b; Rubin & Donkin, 2024).

# 4. Areas of Agreement and Disagreement

## Areas of Agreement

Despite their contrasting views on *p*-hacking, there are several areas of agreement between the error statistical and formal inference approaches that are worth highlighting. First, they both acknowledge that their diverging views reflect different philosophies of science (e.g., Mayo, 1996, p. 255; Mayo & Cox, 2010, p. 267; Rubin, 2024b, pp. 46–47; Rubin, 2025). As Mayo and Cox (2010) explained:

> The general issue is whether the evidential bearing of data $y$ on an inference or hypothesis $H_0$ is altered when $H_0$ has been either constructed or selected for testing in such a way as to result in a specific observed relation between $H_0$ and $y$, whether that is agreement or disagreement. Those who favor logical approaches to confirmation say no (e.g., Mill 1888,

> Keynes, 1921), whereas those closer to an error-statistical conception say yes (Whewell (1847), Peirce (1931–5)). (p. 267)

Error statisticians also acknowledge the logical and mathematical arguments that support the formal inference approach, although they find these arguments wanting. For example, Mayo and Cox (2010) explained that, "to the extent that *p* is viewed as an aspect of the logical or mathematical relation between the data and the probability model, … preliminary choices [based on an inspection of the data] are irrelevant" (p. 266). Similarly, Mayo (1996) explained that, "if confirmation is strictly a logical function between evidence (or statements of evidence) and hypotheses, when or how hypotheses are constructed will be irrelevant" (p. 255).

Finally, the two approaches agree that a researcher's personal biases may affect the specification of a significance test without invalidating its Type I error rate. As Mayo (1996) explained:

> It does not matter that test specifications might reflect the beliefs, biases, or hopes of the researcher. Perhaps the reason for selecting an insensitive test is your personal desire to find no increased risk, or perhaps it is due to economic or ethical factors. Those factors are entirely irrelevant to scrutinizing what the data do and do not say. They pose no obstacle to my scrutinizing any claims you might make based on the tests, nor to my criticizing your choice of test as inappropriate for given learning goals. (p. 409; see also pp. 148, 263)
>
> The latitude that exists in the choice of test does not prevent the determination of what a given result does and does not say. The error probabilistic properties of a test procedure — *however that test was chosen* — allows for an objective interpretation of the results. (p. 406)

## Areas of Disagreement

Despite agreeing that personally biased test specification does not prevent an objective interpretation of test results, the two approaches disagree about the impact of undisclosed result-dependent selective reporting (i.e., *p*-hacking). According to the error statistical approach, *p*-hacking invalidates significance testing because it represents an unrecognized part of the relevant ("actual") test procedure. In contrast, in the formal inference approach, *p*-hacking does not represent part of the relevant (formally reported) test procedure. Consequently, the formal inference approach regards *p*-hacking in the same way that the error statistical approach regards personally biased test specification: Bias may occur in both cases, but it does not prevent either (a) public scrutiny of the formally reported rationale for a specified test (Rubin, 2022, pp. 539–542; Rubin & Donkin, 2024, pp. 2024, 2036–2037) or (b) an objective interpretation of the test's results relative to a formally specified sampling distribution

From a broader perspective, the two approaches disagree about where to draw the line between the *context of discovery* and the *context of justification* (Reichenbach, 1938). The context of discovery includes a researcher's "processes of thinking in their actual occurrence," whereas the context of justification includes the "rational reconstruction" or "logical substitute" of these subjective thought processes (Reichenbach, 1938, pp. 6–7). The error statistical approach argues that the construction and selection of hypotheses and data occur in the context of justification. In contrast, the formal inference approach follows Popper's (1983) view that "the factual, psychological, and historical question, 'How do we come by our theories?' [or statistical hypotheses], though it may be fascinating, is irrelevant to the logical, methodological, and epistemological question of validity" (p. 36; see also Popper, 2002, pp. 7–8, 22). From this perspective, it does not matter that statistical hypotheses may be secretly selected because their test results are psychologically "desired" by researchers (e.g., Nagy et al., 2025, p. 2; Simmons et al., 2011, p. 1359; Stefan & Schönbrodt, 2023, p. 2). What matters is a public critical evaluation of the formally reported test procedures and inferences (see also Rubin, 2025, p. 5).

Mayo (1996) recognized these diverging views when defining the "actual" test procedure in the context of an example in which a researcher conducted multiple tests and then searched for a significant result to report:

> It may be objected that … I am taking sides in favor of one description of the "actual" test procedure — one that takes into account the fact that searching has occurred. I am, but maintain that this aspect of the procedure cannot be ignored given the aim of the statistical significance test chosen. Remember, I am distinguishing the appropriateness of the test chosen (for a given inquiry) from the error probabilities, *given* that that test is chosen. (p. 304)

Mayo's (1996) position makes sense from an error statistical perspective because ignoring the undisclosed searching for significance misspecifies the "actual" test procedure. However, in the formal inference approach, "the aim of the statistical test chosen" is to tentatively rule out random sampling error "*given* that that test is chosen" and "*however that test was chosen"* (Mayo, 1996, pp. 304, 406). From this perspective, it is legitimate to define the test procedure solely in terms of its formal, publicly reported specification and independent from any personally biased, result-dependent selection process that influenced that specification (see also Mayo, 1996, p. 255; Mayo & Cox, 2010, p. 266).

Mayo and Cox (2010) disagree, arguing that, "by allowing the result to influence the choice of specification, one is altering the procedure giving rise to the *p*-value" (p. 271). Again, they are correct from an error statistical perspective because an error statistical test includes the process by which its hypothesis and data were constructed and selected (Mayo, 1996, pp. 206–207, 298). Consequently, allowing a test's result to influence this process will alter the "actual" test procedure. However, from a formal inference perspective, a significance test and its *p*-value are only conditioned on the formally reported test procedure and not on the "factual, psychological, and historical" process that led up to that procedure in the context of discovery (Popper, 1983, p. 36). Consequently, changing the process by which a test procedure is specified does not alter the procedure itself.

Mayo (2018) claimed that "selection effects alter the outcomes in the sample space, showing up in altered error probabilities" (p. 286). Again, this claim is correct if we follow the error statistical approach and extend the sample space to include outcomes that would arise during hypothetical repetitions of the selection process. However, it does not apply in the formal inference approach, which limits the relevant sample space to outcomes under the selected statistical null hypothesis. For example, a biasing selection effect will not alter the Type I error rate for an inference about $H_{0,13}$ when that inference is limited to potential outcomes in $H_{0,13}$'s sample space, even if $p_{13}$;$H_{0,13}$ has been *p*-hacked from a set of unreported tests $p_1$;$H_{0,1}$, … $p_k$;$H_{0,k}$ (Rubin, 2024b, p. 55).

Spanos and Mayo (2015) argued that "the discrepancy between actual and nominal error probabilities stems from evaluating the nominal error probabilities using the wrong sampling distribution" (p. 3546; see also Figure 1). Certainly, from an error statistical perspective, the sampling distribution under $H_{0,13}$ does not reflect the "actual" test procedure when $p_{13};H_{0,13}$ has been *p*-hacked from $p_1;H_{0,1}$, $p_2;H_{0,2}$, $p_3;H_{0,3}$, … $p_k;H_{0,k}$. However, from a formal inference perspective, the sampling distribution under the intersection null hypothesis $H_{0,1} \cap H_{0,2} \cap H_{0,3} \ldots \cap H_{0,k}$ is the "wrong" distribution because it is incommensurate and inconsistent with the researcher's formally reported inference, which is only about the individual hypothesis $H_{0,13}$ (Rubin, 2024a, p. 3; Rubin, 2025, pp. 13–14).

Finally, Mayo (2018) returns to the issue of well-testedness or severity:

> For the severe tester, outputting $H_{13}$, ignoring the non-significant others, renders $H_{13}$ poorly tested. You might say, but look there's the hypothesis $H_{13}$, and data $x$ — shouldn't it speak for itself? No. That's just the evidential-relationship or logicist in you coming out. (p. 275)

Again, however, in the formal inference approach, the well-testedness of $H_{0,13}$ depends on the individual test of $H_{0,13}$, not the union-intersection test of $H_{0,1} \cap H_{0,2} \cap H_{0,3} \ldots \cap H_{0,k}$. In this respect, formal inferentialists are not afraid to let their inner logicist come out! Indeed, they would view it as illogical to use the "actual" familywise error rate to license an individual inference about $H_{0,13}$ because this approach commits the fallacy of division (Selvin, 1958; Waller, 2018 for examples of this fallacy in real research studies, see Rubin, 2024a, pp. 5–6).

In summary, disagreements between the error statistical and formal inference approaches stem from their different conceptualizations of test procedures, the context of justification, sample spaces, and sampling distributions. The error statistical approach assumes that *p*-hacking (a) constitutes part of the "actual" test procedure, (b) occurs in the context of justification, and (c) alters the relevant sample space and sampling distribution. In contrast, the formal inference approach assumes that *p*-hacking (a) is not part of the selected, formally reported test procedure, (b) occurs in the context of discovery and, consequently, (c) does not alter the relevant hypothetical sample space or sampling distribution (Rubin, 2025, pp. 4–5, 9–10).

# 5. The Minimal Severity Requirement

According to the error statistical approach's minimal severity requirement, it is important to adequately check a test's assumptions in order to provisionally rule out systematic error, because systematic error can cause a significant discrepancy between "actual" and nominal error probabilities that invalidates nominal error probabilities (Mayo & Spanos, 2011, p. 190; Spanos, 2010, p. 219; Spanos & Mayo, 2015, pp. 3541, 3543). In the formal inference approach, however, nominal Type I error rates are not invalidated by systematic error because they refer to a *hypothetical* situation in which there is no systematic error, all test assumptions are met, and the only source of decision-making error is random sampling error (Fisher, 1956, pp. 40, 44, 77–78, 82; Fisher, 1959, pp. 21–22; Kass, 2011, pp. 5–6, 7; Meehl, 1990, p. 111; Neyman, 1955, p. 17; Neyman & Pearson, 1928, pp. 177, 232; Rubin, 2024b, p. 49; see also Popper, 1983, p. 313). Hence, it is not necessary to meet the minimal severity requirement in order to obtain valid Type I error rates in the formal inference approach. This point can be illustrated with respect to (a) implicatory assumptions and (b) the Texas sharpshooter fallacy.

## Implicatory Assumptions

In the formal inference approach, a Type I error rate refers to a hypothetical situation in which random sampling error is the only source of decision-making error. This hypothetical situation can be described as an "*implicationary* or *i-assumption*" that is made during testing in order to draw out certain logical implications (Mayo, 2018, pp. 109, 167). According to Mayo (2018), "the howler [mistake] occurs when a test hypothesis that serves merely as an i-assumption is purported to be an actual assumption, needed for the inference to go through" (p. 167; see also Cortina & Dunlap, 1997, pp. 166−167). The formal inference approach agrees. However, it applies the same reasoning to not only the test hypothesis but also the test assumptions (see also Hennig, 2010, p. 47; Hennig, 2023, pp. 24, 38). Hence, the formal inference approach does not require a test's "actual" assumptions to be checked and affirmed as "approximately true" in order for a statistical inference to go through (Mayo, 2011, p. 96; Mayo & Spanos, 2011, p. 189; cf. Hennig, 2023, p. 38). Instead, during testing, all test assumptions are provisionally accepted as i-assumptions. In this case, failure to probe for systematic error in a test's "actual" assumptions does not invalidate that test's nominal error rate because the associated inferential argument is based on unproblematic i-assumptions.

The relation between "actual" and implicatory assumptions is similar to that between the soundness and validity of a logical argument. For example, consider the argument that "all cats are gray, and *x* is a cat; therefore, *x* is gray." This argument is logically valid, even though it is unsound because its major premise is not "approximately true" — not all cats are gray; not even most cats are gray (Fisher, 1959, pp. 22−23; Mayo, 2018, p. 60). Furthermore, the argument remains valid even though other major premises are available that *are* approximately true (e.g., "all cats have whiskers"). Similarly, in the formal inference approach, it remains valid to use a nominal Type I error rate to license a statistical inference even when the associated test's "actual" assumptions are not approximately true. In other words, it remains valid to make the suppositional, subjunctive conditional argument that, *if* a test's assumptions *were* true and random sampling error was the only source of decision-making error, *then* the nominal Type I error rate would represent the maximum frequency with which the null hypothesis would be incorrectly rejected (Neyman, 1950, p. 289; Neyman, 1955, p. 17; see also Cortina & Dunlap, 1997, pp. 166–167; Fisher, 1959, pp. 21–22; Hennig, 2023, pp. 21, 23–24; Kass, 2011, pp. 5–6, 7; Mayo, 2018, pp. 426–427). In this case, the nominal Type I error rate represents a *hypothetical* error rate rather than an "actual" error rate, and "we speak of what *would* be inferred if our assumptions *were* to hold" (Kass, 2011, p. 7; see also Fisher, 1956, pp. 44, 77).[10]

The use of i-assumptions during hypothesis testing is consistent with Popper, who argued that a test's background assumptions (initial conditions and auxiliary hypotheses) must be accepted as "unproblematic" during the test in order to force the test hypothesis into logical isolation (Popper, 1983, pp. 186, 244; Popper, 1994, p. 160; Popper, 2002, pp. 197, 260; see also Lakatos, 1978, pp. 32, 42; Rubin, 2025, p. 12, Footnote 7). Of course, a test's "actual" assumptions may be challenged before or after the test takes place (Lakatos, 1978, p. 158; Popper, 1994, p. 160), and if they are judged to be inadequate, then an alternative test may be considered (Popper, 2002, p. 86). However, in the formal inference approach, the identification of a more adequate test does not invalidate a *statistical* inference based on the original (inadequate) test because the inferential argument for that test is based on hypothetical i-assumptions, not real-world "actual" assumptions. "2 + 2 = 4," and this mathematical equation remains valid even if one of the "2"s later turns out to be a "3."

## The Texas Sharpshooter Fallacy

Mayo also used the Texas sharpshooter fallacy to illustrate the error statistical concern about violating the minimal severity requirement (Mayo, 1996, pp. 201–203; Mayo, 2018, pp. 5, 19, 276–277; Mayo & Cox, 2010, p. 271). In this scenario, an individual with poor shooting skills fires several shots at a barn wall and then paints a target around a random cluster of their bullet holes. Later on, they show these "bullseyes" to their friend as evidence of their "excellent shooting skills." The minimal severity requirement is not met in this scenario because there has been no attempt to rule out the sharpshooter's fraudulent behind-the-scenes behavior as one of the ways in which their claim may be false (Mayo, 2018, p. 19). Consequently, we are left with "bad evidence, no test (BENT)" (Mayo, 2018, p. 5).

Again, however, from a formal inference perspective, it is not necessary to meet the minimal severity requirement in order to report valid hypothetical probability statements. For example, it remains valid for the sharpshooter to compute the probability that their shots would hit the painted target in repetitions of their formally reported, hypothetical, test procedure (i.e., the procedure that they reported to their friend; see also Mayo, 2018, p. 19). In this case, the relevant probability is a formal mathematical probability about the occurrence of a hypothetical event based on chance alone, and so it is not invalidated by the presence of fraud in the real world (Fisher, 1956, pp. 40, 44, 77; Fisher, 1959; Neyman, 1955, p. 17; Neyman & Pearson, 1928, pp. 176, 232; cf. Mayo, 2018, p. 19).

To be clear, I am not arguing that the sharpshooter's fraudulent behavior is unproblematic. It certainly invalidates the *substantive* (nonstatistical) inference about their "excellent shooting skills." I am only arguing that, during significance testing, Type I error rates license hypothetical, test-specific, *statistical* inferences given random sampling error alone, not real-world, general, *substantive* inferences given systematic error (Meehl, 1990, p. 111; Meehl, 1997, p. 401; Neyman, 1950, p. 290). Consequently, we should not expect Type I error rates to be "altered," "invalidated," or "vitiated" by the presence of systematic error in the real world, including error caused by fraud.

By analogy, we would not expect a weighing scale's random measurement error to be altered or invalidated when a person lies about their weight or when the scale systematically underestimates weight. These systematic errors will certainly invalidate substantive inferences about the person's weight, but they will not invalidate statistical inferences based on the scale's random measurement error.

## 6. The Nondisclosure of Null Results

*p*-hacking results in the nondisclosure of null (nonsignificant) results. Again, this nondisclosure only inflates Type I error rates in the error statistical approach. Nonetheless, it may also be problematic in the formal inference approach, depending on how we conceptualize null results and their impact on substantive inferences.

To illustrate, consider a researcher who reports the substantive inference that "eating jelly beans causes acne" (Munroe, 2011). Imagine that they base this substantive inference on a single statistical inference: "Participants who ate green jelly beans had significantly more acne than those in a control group, $t(326) = 2.63$, $p = 0.009$." However, unbeknownst to readers, the researcher does not report the results of 19 other tests that found no significant results using 19 other colors of jelly beans ($p$s > 0.05).

From an error statistical perspective, the "actual" familywise error rate for the statistical inference (0.642) is inflated above the conventional nominal error rate (0.05) because the "actual" test procedure includes 20 tests, not 1 test. In contrast, from a formal inference perspective, the nominal Type I error rate of 0.05 is valid because the reported statistical inference is an individual inference about green jelly beans, not a union-intersection inference about green, red, blue, and other colors of jelly beans (for a discussion, see Rubin, 2021b, pp. 10978–10983). But does the researcher's failure to report the 19 null results bias their broader, nonstatistical, *substantive* inference that "eating jelly beans causes acne"? The answer depends on how we conceptualize null results and how they relate to the substantive inference.

First, null results may be regarded as providing *evidence of absence* because we tentatively "accept" null hypotheses following high-powered tests and severe inferences (Neyman, 1950, pp. 259–260; e.g., Mayo, 2018, p. 150; Mayo & Cox, 2010, p. 256; Mayo & Spanos, 2006, p. 339; Mayo & Spanos, 2011, p. 177). In this case, nondisclosed null results may bias substantive inferences. For example, in the current scenario, 19 results indicate that jelly beans *do not* cause acne, and only one result indicates that they *do*! Hence, failing to report the 19 null results is problematic because it hides relevant evidence and biases the evidential support for the substantive inference.

Second, null results may be regarded as the *absence of evidence* (Aczel et al., 2018; Altman & Bland, 1995; Murphy et al., 2025; Rubin & Donkin, 2024, p. 2036) because we merely "fail to reject" null hypotheses following low-powered tests, Fisherian tests, and/or nonsevere inferences (Fisher, 1956, p. 45; Mayo & Spanos, 2006, pp. 338–339; Mayo & Spanos, 2011, p. 176; Neyman, 1955, p. 41). In this case, null results do not impact substantive inferences because, in the absence of a meta-analysis, they do not provide any evidence one way or the other. For example, in the current scenario, the single significant result for green jelly beans supports the substantive claim that "eating jelly beans causes acne," whereas the 19 null results do not count as evidence either for or against this claim and so can be ignored.[11] From this perspective, it is more problematic for researchers to omit tests that yield *significant* results that *contradict* their substantive inference (i.e., significant negative results rather than null results). For example, it would be problematic if the researcher failed to report a result showing that eating *red* jelly beans significantly *reduced* acne relative to a control group. In this case, they would be omitting relevant evidence that contradicted the substantive claim that "eating jelly beans causes acne" (Rubin, 2020, p. 384).

Finally, even if null results are regarded as providing evidence of absence, their nondisclosure will only bias a substantive inference when they are *theoretically relevant* to that inference (Rubin, 2020, p. 378; Rubin, 2022, p. 548). For example, failing to report the severe, high-powered, statistical inference that "there was no significant difference in acne between participants who ate M&Ms and those in a control group, $t(326) = 0.33$, $p = 0.742$" will not bias the substantive claim that "eating jelly beans causes acne" because M&Ms are not jelly beans (Rubin, 2021b, p. 10988). Hence, it is important to distinguish between *biased* selective reporting, which is potentially problematic, and *unbiased* selective reporting, which is an essential part of effective scientific communication (Rubin, 2020, p. 383). A theory-centric critical rational discussion among researchers is required to make this distinction (Popper, 1966, pp. 218–219, 229–231; Popper, 1974, p. 82; Popper, 1994, pp. 93, 159–160; Rubin, 2025, p. 11).

In summary, nondisclosed null results only bias substantive inferences when they are regarded as being both (a) evidence of absence and (b) theoretically relevant. How can we identify these evidentially and theoretically important null results? Preregistration represents one approach. However, preregistration cannot be used to identify nondisclosed null results in the case of either exploratory analyses or unreported deviations from preregistered analyses. For example, a researcher may try out two deviations from a preregistered analysis protocol and then only report the deviation that yields a significant result. In this case, cross-referencing between the final research report and the preregistered protocol will not help to identify the undisclosed null result.

A more comprehensive approach to identifying theoretically important null results is to engage in an "imaginative" critical rational discussion of relevant theory, evidence, and background knowledge (Popper, 1974, p. 148; Popper, 2002, p. 466; Popper, 1966, 1994). As Kerr (1998) asked, if an original hypothesis is not reported, but it "had a sufficient rationale (theoretical, empirical, or even intuitive) to recommend itself to one researcher, why would it not also occur to others?" (p. 208; see also Rubin, 2020, p. 378; Rubin, 2022, p. 548). For example, in the current scenario, we might reasonably expect reviewers and other readers to ask why the researcher only used *green* jelly beans to test the substantive hypothesis that "eating jelly beans causes acne" and whether the researcher tested any other colors of jelly beans (Wells & Windschitl, 1999). Furthermore, researchers' anticipation of this critical discussion of their work helps to constrain their "researcher degrees of freedom" (Simmons et al., 2011, p. 1359) and limit an "anything goes" approach (Hennig, 2023, pp. 17, 19).[12] Finally, this "imaginative criticism" (Popper, 1974, p. 148; Popper, 2002, p. 466) is more useful than preregistration because it will reveal not only relevant hypothesis tests that were conducted and suppressed, but also relevant hypothesis tests that were not even considered by the original researcher.

# 7. Conclusions and Implications

## Conclusions

I have argued that *p*-hacking inflates Type I error rates in the error statistical approach but not in the formal inference approach. In the error statistical approach, the "actual" test procedure determines the relevant error rate. In contrast, in the formal inference approach, the formally reported statistical inference determines the relevant error rate. *p*-hacking is part of the "actual" test procedure, but it is not part of the hypothetical test procedure based on a formally reported inference. Consequently, *p*-hacking inflates the relevant error rate in the error statistical approach but not in the formal inference approach.

Importantly, from a formal inference perspective, it is inappropriate to use the "actual" familywise error rate to license inferences about selectively reported hypotheses. Logically, the "actual" familywise error rate only licenses an inference about the entire family of reported and unreported hypotheses that are tested in the "actual" test procedure. It does not license inferences about selectively reported hypotheses, which are licensed by their own nominal error rates. Certainly, *p*-hacking influences the *selection* of reported significance tests. However, this selection process is not *part of* those tests, and so *p*-hacking does not influence the associated nominal Type I error rates. Hence, in the formal inference approach, "computed" nominal Type I error rates can be taken at face value, barring any logical or mathematical errors (Neyman, 1950, p. 289; Rubin, 2024b, p. 53; Rubin, 2025, p. 14).

In the error statistical approach, failure to meet the minimal severity requirement leaves us with "bad evidence, no test (BENT)" (Mayo, 2018, p. 5). In contrast, in the formal inference approach, failure to rule out ways in which a substantive inference could be false may result in a "bad" *substantive* inference without invalidating a corresponding *statistical* inference, which is based on tentatively accepted i-assumptions. Of course, a researcher's "actual" test assumptions may be judged to be inadequate. In this case, they should use a different test with more adequate assumptions. However, doing so only has the potential to reduce systematic error relative to a specified research goal. It will not result in a more valid Type I error rate because Type I error rates do not represent systematic error in a test's selection, specification, or assumptions. They only represent random error *given* a test's selection, specification, and assumptions.

Finally, nondisclosed null results inflate Type I error rates in the error statistical approach but not in the formal inference approach. Nonetheless, nondisclosed null results may be problematic in the formal inference approach if they are regarded as being both (a) evidence of absence and (b) theoretically relevant. In this case, although unreported null results do not inflate Type I error rates, they may bias substantive inferences. However, this substantive bias can be identified and addressed through an imaginative, theory-centric, critical rational discussion among peers (Popper, 1966, 1994; Rubin, 2025). This social epistemological approach to bias (e.g., O'Rourke-Friel, 2025, p. 7) stands in contrast to the error statistical approach's more individualistic statistical approach (Rubin, 2025, p. 11).

## Implications

The contrast between the error statistical and formal inference approaches has implications for the way in which we conceptualize *p*-hacking. *p*-hacking is often described as a "questionable" research practice (John et al., 2012). The current analysis supports this equivocal description as opposed to stricter characterizations that equate *p*-hacking with fraud and unethical research behavior (e.g., Miller et al., 2025; Pickett & Roche, 2018). In particular, *p*-hacking may be viewed as being more or less problematic for Type I error rates depending on one's philosophy of significance testing. Accordingly, it would be helpful for discussions about the potential dangers of *p*-hacking to be more clearly situated within relevant philosophies of significance testing in order to highlight otherwise implicit assumptions.

A similar recommendation applies to demonstrations and simulations showing that *p*-hacking inflates Type I error rates. This work tends to adopt an error statistical perspective by contrasting nominal and "actual" error rates (e.g., Simmons et al., 2011, p. 1359). However, it should be acknowledged that, during *p*-hacking, (a) researchers cannot report union-intersection inferences about "actual" intersection null hypotheses, (b) "actual" familywise error rates cannot be computed due to the indeterminate number of tests in a hypothetical long run of sampling, and, even if they could be computed, (c) "actual" familywise error rates do not license researchers' selectively reported individual inferences (Rubin, 2024b, pp. 58, 63).

The contrast between the error statistical and formal inference approaches also has implications for interventions that are intended to reduce *p*-hacking. In particular, from an error statistical perspective, preregistration provides a useful method of limiting *p*-hacking, Type I error rate inflation, and the nondisclosure of null results (Lakens, 2019; Mayo, 2018, pp. 106–107, 439). However, from a formal inference perspective, *p*-hacking does not alter relevant (nominal) Type I error rates (Rubin, 2024b, 2025). In addition, nondisclosed null results may (a) represent the absence of evidence, (b) be theoretically irrelevant, and (c) be identified following a critical rational discussion of relevant theory and background knowledge (Rubin, 2020, p. 378; Rubin, 2022, p. 548; Rubin & Donkin, 2024, p. 2036). Hence, the formal inference approach does not provide a rationale for preregistration based on either Type I error rate inflation or the nondisclosure of null results.

Turning to the replication crisis, error statisticians have argued that "inflated error rates [are] at the heart of obstacles to replication" (Mayo, 2025, p. 1) and that "much of the handwringing about irreproducibility is the result of wearing blinders as to the construction and selection of both hypotheses and data" (Mayo, 2018, p. 49). In contrast, from a formal inference perspective, researchers who *p*-hack do not report statistical inferences that have inflated Type I error rates because they do not report union-intersection inferences about "actual" intersection null hypotheses (Rubin, 2024b). Instead, they selectively report individual inferences that have valid nominal error rates.

To be clear, the formal inference approach attributes replication failures to *both* statistical Type I errors *and* substantive systematic errors. However, it argues that, by definition, Type I error rates are fixed at nominal levels conditioned on formally specified statistical models. In contrast, unrecognised systematic errors may occur more frequently than expected, which may help to explain lower than expected replication rates (Rubin, 2024b, pp. 63–64; Tyner et al., 2026, p. 147).

Of course, it is important to undertake rigorous auditing in original studies to try to identify systematic errors (Mayo, 2018, pp. 100, 441–442). However, auditing can only examine *known unknowns* — the potential systematic errors that researchers have thought to investigate. By definition, even the most rigorous auditing cannot identify *unknown unknowns* — the potential systematic errors that researchers have not yet considered. Unlike known unknowns, unknown unknowns cannot be tentatively ruled out in original studies, and they may go on to represent hidden moderators that cause replication failures (Rubin, 2024b, pp. 63–64; Tyner et al., 2026, p. 147; for examples, see Firestein, 2016).

Hence, from a formal inference perspective, the replication crisis is not the result of *p*-hacking inflating Type I error rates. Certainly, Type I errors will result in *some* replication failures, but no more than expected based on the nominal conventional alpha level. The replication crisis is also not the result of poorly audited test assumptions. More rigorous auditing will help to tentatively rule out *some* systematic errors in original studies. However, by definition, it will be unable to identify unrecognized systematic errors (unknown unknowns) that go on to cause replication failures via hidden moderators.

Instead, the replication crisis is a result of our underestimation of the extent and impact of unrecognized systematic errors. We end up being surprised by the extent of our own unanticipated ignorance! From this perspective, the solution to the replication crisis will not be found in new statistical or methodological approaches. Instead, it requires a more fundamental epistemological shift in our research culture that produces a better acknowledgment "of the finitude and fallibility of our knowledge, and of the infinity of our ignorance" (Popper, 1994, p. 123). In particular, we must accept that our carefully constructed experiments, methods, and models also represent theories and, as such, they and their assumptions are fallible and often misleading (Lakatos, 1978, p. 44).

## Endnotes

1. Following the error statistical approach, a semicolon ";" is used to indicate that a *p*-value is computed *assuming* that a specific null hypothesis is true (e.g., "$p_1;H_{0,1}$"; Mayo & Morey, 2017, p. 12; Mayo & Spanos, 2006, p. 331; Mayo & Spanos, 2011, p. 169).
2. The formula for computing the familywise error rate assumes that test variables are independent. If this assumption is not met and variables are positively correlated with one another, then the "actual" familywise error rate will be less than $1 - (1 - \alpha)^k$. Hence, this formula provides an upper bound or conservative, worst-case value (Mayo, 2018, p. 275; Rubin, 2021, p. 10974, Footnote 3). For simplicity, I equate this upper bound with the "actual" familywise error rate throughout.
3. The word *actual* is placed in scare quotes throughout because, from a formal inference perspective, the "actual" (i.e., inferentially relevant) test procedure and Type I error rate are hypothetical and formally specified (Rubin, 2025, pp. 13–14).
4. Another method of correcting for multiple testing is based on the false discovery rate (FDR; Benjamini & Hochberg, 1995, p. 290). The familywise error rate is the probability of rejecting at least one constituent null hypothesis (e.g., $H_{0,1}$) in an intersection null hypothesis (e.g., $H_{0,1} \cap H_{0,2}$) when all constituent hypotheses are true. In contrast, the FDR is the expected proportion of incorrectly rejected null hypotheses among *all* (true *and* false) rejected null hypotheses. Hence, unlike the familywise error rate, the FDR assumes that some of the constituent null hypotheses in an intersection null hypothesis (e.g., $H_{0,1}$) may be *correctly* rejected because they are *false* (Rubin, 2021b, p. 10975). Consequently, from a formal inference perspective, the FDR does not represent a Type I error rate because Type I error rates are computed under the assumption that the associated null hypothesis is true (Neyman, 1952, p. 57), and the FDR assumes that the intersection null hypothesis (e.g., $H_{0,1} \cap H_{0,2}$) may be false because one or more of its constituent hypotheses may be false (e.g., $H_{0,1}$).

5. A "crud" effect refers to the "crud factor," which is an "epistemological concept that, in correlational research, all variables are connected through causal structures, which result in real nonzero correlations between all variables in any given data set" (Orben & Lakens, 2020, p. 241). It is important to stress that "*the crud factor does not refer to statistical error, whether of the first or the second kind*" (Meehl, 1997, p. 402, emphasis in original). In other words, the crud factor does not refer to either Type I or Type II errors.
6. In response to Rubin (2021b), Goeman (2022) argued that strong control over the familywise error rate (e.g., via a Bonferroni correction) allows us to make inferences about not only (a) a superordinate intersection null hypothesis (e.g., $H_{0,1} \cap H_{0,2} \cap H_{0,3}$), but also (b) subordinate intersection null hypotheses (i.e., $H_{0,1} \cap H_{0,2}$, $H_{0,2} \cap H_{0,3}$, and $H_{0,1} \cap H_{0,3}$) and (c) individual null hypotheses (e.g., $H_{0,1}$, $H_{0,2}$, and $H_{0,3}$). However, Goeman's argument confuses (a) constituent null hypotheses that are treated as being logically exchangeable parts of the same superordinate intersection null hypothesis with (b) subordinate intersection and individual null hypotheses that are treated as being logically distinct and nonexchangeable with one another (García-Pérez, 2023, p. 2; Rubin, 2021b, pp. 10978–10982; Rubin, 2024a, p. 2). For example, the familywise error rate for a union-intersection inference about $H_{0,1} \cap H_{0,2} \cap H_{0,3}$ only licenses a decision about $H_{0,1} \cap H_{0,2} \cap H_{0,3}$. It would be a fallacy of division (Waller, 2018) to argue that it also licenses decisions about subordinate intersection null hypotheses, such as $H_{0,1} \cap H_{0,2}$, and individual null hypotheses, such as $H_{0,1}$, which are licensed by their own error rates. Furthermore, in the formal inference approach, researchers should not be interested in the error rate for $H_{0,1} \cap H_{0,2} \cap H_{0,3}$ unless they report a union-intersection inference about this hypothesis. Customized, inference-specific error control at the level of formally reported intersection and/or individual hypotheses provides a more logically consistent and statistically powerful approach, and it makes strong error control over superordinate intersection null hypotheses irrelevant (see also Rubin, 2021b, pp. 10990–10992).
7. Throughout this article, I assume that researchers tend to report individual inferences rather than union-intersection or intersection-union inferences because individual inferences are the most common type of statistical inference (García-Pérez, 2023, p. 4). However, the same formal inference principles apply to subordinate intersection null hypotheses (e.g., $H_{0,1} \cap H_{0,2}$) that have been selectively reported from an "actual" superordinate intersection null hypothesis (e.g., $H_{0,1} \cap H_{0,2} \cap H_{0,3}$). For a discussion and illustration, see Rubin (2024b, pp. 59–60).
8. The formal inference approach can also be applied to the case of undisclosed optional stopping (Rubin, 2024b, pp. 61–62). In this case, the concern is that the alpha level needs to be adjusted to account for the multiple testing of an intersection null hypothesis that may be rejected following a significant result on at least one of the constituent tests that have been conducted during the course of the optional stopping process (e.g., $p_1;H_{0,1}$, $p_2;H_{0,2}$, and $p_3;H_{0,3}$). From a formal inference perspective, this concern is only warranted if researchers report a formal inference about the intersection null hypothesis that comprises the associated constituent null hypotheses (i.e., $H_{0,1} \cap H_{0,2} \cap H_{0,3}$). By definition, researchers cannot report this union-intersection inference during *undisclosed* optional stopping (i.e., *p*-hacking) because they do not report the interim tests (e.g., $p_1;H_{0,1}$ and $p_2;H_{0,2}$). Instead, they report an *individual* inference about the individual null hypothesis that is tested by their final, formally reported test (i.e., $H_{0,3}$). As discussed previously, no alpha adjustment is required in the case of this individual inference.

9. Mayo (2018, p. 14) proposed an "argument from coincidence" as a "positive counterpart" to the minimal severity requirement. As Mayo (2010) explained, "the most familiar canonical exemplar of an argument from coincidence is statistical null hypothesis testing … [Here,] the null hypothesis, $H_0$, asserts that the effect or observed agreement is 'due to coincidence'" (p. 85; see also Fisher, 1956, p. 35). When testing an intersection null hypothesis, the "coincidence" is that at least one test would yield a significant result when all constituent null hypotheses are true. If the familywise error rate is suitably low, then there is a low probability of encountering this coincidence. In this case, we can say that the union-intersection test has a high "capacity" to detect an error, and we are licensed to argue against this "coincidence" to the absence of error in the case at hand. Crucially, however, when testing an intersection null hypothesis, the case at hand is an *intersection* null hypothesis, not an *individual* hypothesis.
10. For Mayo, an inductive or ampliative inference is one that has been detached from its inferential argument (e.g., "*x* is gray" in the case of "all cats are gray, and *x* is a cat; therefore, *x* is gray"; Mayo, 2010, p. 82; Mayo, 2011, pp. 84–85; Mayo, 2018, p. 65). To be warranted in the error statistical approach, a statistical inference must be detached after the argument's premises have been adequately checked and affirmed as being "approximately true" (Mayo, 2011, pp. 85, 96; Mayo & Spanos, 2011, pp. 189, 193; see also Mayo, 1996, p. 435; Mayo, 2018, pp. 84, 441–442). In contrast, in the formal inference approach, a *statistical* inference cannot be detached from its inferential argument because it is subjunctively conditioned on a test's suppositional i-assumptions (Kass, 2011, pp. 5–6, 7; see also Fisher, 1956, p. 44; Hennig, 2023, p. 24; Mayo, 2018, p. 427). Only a nonstatistical *substantive* inference can be detached as a more general standalone conclusion. Nonetheless, following Fisher (1959, p. 21), a statistical inference can be regarded as a suppositional "as if" inductive inference. Specifically, it can be argued that *if* a test's i-assumptions *were* true, *then* a significant result *would* license the provisional inference that the study's current sample does not belong to a specified hypothetical infinite null population that contains no known relevant subsets (Fisher, 1956, pp. 32–33, 57, 109; Fisher, 1959, pp. 21–24). Note that, despite its name, this "inductive reasoning" is consistent with Popper's deductive approach because it is conceptualized as a conditional "statement of uncertainty" about the observed result given our specified ignorance (Fisher, 1956, pp. 29, 57, 108–110; Fisher, 1959, p. 22; see also Greenland, 1998, pp. 545–546; Senn, 1991, p. 1683) rather than an enumerative induction about future results during "repeated sampling from the same population" (Fisher, 1956, p. 77).
11. The difference between a significant result and a nonsignificant result may not itself be significant (Gelman & Stern, 2006). Hence, at this stage, we cannot draw any conclusions about the potential moderating effect of jelly bean color.
12. In order to avoid questions about why they only tested *green* jelly beans, a researcher may deliberately omit jelly bean color from the specification of their statistical inference. In the formal inference approach, this personally biased respecification does not impact the validity of the associated Type I error rate, which is subjunctively conditioned on the researcher's formally reported inference. Like all statistical inferences, the researcher's respecified inference may include irrelevant variables and/or exclude relevant variables (Rubin, 2025, p.

7, Footnote 5). Hence, when interpreting statistical inferences, it is important to appreciate "that the population in question is hypothetical, that it could be defined in many ways, and that the first to come to mind may be quite misleading" (Fisher, 1956, p. 78). In the present case, future tests may confirm that color is an irrelevant variable. Alternatively, future tests may reveal color to be a previously hidden moderator that had caused a systematic error in the substantive inference that "eating jelly beans causes acne" (i.e., only eating *green* jelly beans causes acne). However, this revelation would not invalidate the researcher's original statistical inference, which requires us to adopt a potentially counterfactual "postulate of ignorance" (i-assumption) about the presence of relevant subsets (e.g., *green* jelly beans) within a hypothetical infinite population (i.e., "jelly beans"; Fisher, 1959, p. 23). This postulate allows us to construe the current sample as a random sample and to attribute uncertainty about the statistical inference to random sampling error independent from systematic error (Fisher, 1956, pp. 32–33, 57, 109; Fisher, 1959, pp. 21–24). Note that, if the discovery of previously hidden moderators and relevant subsets *were* to invalidate prior statistical inferences, then most of our prior statistical inferences would be invalid.

---

*CRediT Statement:* Mark Rubin: Conceptualization, Visualization, Writing — original draft, Writing — reviewing & editing.

*Acknowledgments*: N/A

*Generative AI*: Grammarly, ChatGPT, and Gemini were used to check for typos, grammatical errors, and potential mischaracterisations, as well as to perform a sanity check on the article's arguments. The text and arguments are the author's own work, and they take full responsibility for all of the content.

*Funding:* The author declares no funding sources.

*Conflict of interest:* The author declares no conflict of interest.

*Biography*: Mark Rubin is a professor of psychology at Durham University, UK. For further information about his work in this area, please visit https://sites.google.com/site/markrubinsocialpsychresearch/replication-crisis

*Correspondence:* Correspondence should be addressed to Mark Rubin at the Department of Psychology, Durham University, South Road, Durham, DH1 3LE, UK. E-mail: Mark-Rubin@outlook.com

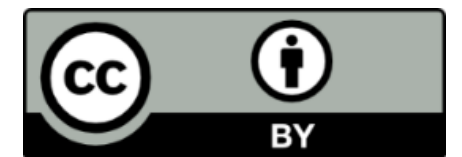